\documentclass[useAMS,usenatbib,12pt]{mn2e}
\usepackage{amsmath}
\usepackage{amssymb}
\usepackage{epsf,graphicx,lscape,subfigure,longtable,txfonts,
}
\usepackage{rotating}

\usepackage[light]{}
\usepackage{multicol}
\usepackage{footnote}
\usepackage{apalike}

\def\kms{\ifmmode{\rm km\,s}^{-1}\, \else km\,s$^{-1}$\,\fi}
\def\mujy{$\mathrm{\muup Jy\,}$}

\def\ltsim{\ifmmode\stackrel{<}{_{\sim}}\else$\stackrel{<}{_{\sim}}$\fi}
\def\gtsim{\ifmmode\stackrel{>}{_{\sim}}\else$\stackrel{>}{_{\sim}}$\fi}
\def\S4195{41.95+575}
\def\S4331{43.31+592}
\def\solmas{$\rm{M_\odot}$}
\def\solmasyr{$\rm{M_\odot yr^{-1}}$}

\def\mdot{$\rm{\dot{M}}\,\,$}

\title[e-MERLIN 21cm radio limits for OB stars in Cyg OB2]{e-MERLIN 21cm constraints on the mass-loss rates of OB stars in Cyg OB2}
\author[J. C. Morford et al.]{J. C. Morford$^1$, D. M. Fenech$^1$, R. K. Prinja$^1$, R. Blomme$^2$, J. A. Yates$^1$\\
$^{1}$Dept. of Physics {\&} Astronomy, University College London, Gower Street, London WC1E 6BT \\
$^{2}$Royal Observatory of Belgium, Ringlaan 3, 1180 Brussel, Belgium \\
}
\begin{document}

\date{}

\pagerange{\pageref{firstpage}--\pageref{lastpage}} \pubyear{}

\maketitle

\label{firstpage}

\begin{abstract}
{We present e-MERLIN 21 cm (L-band) observations of single luminous OB stars in the Cygnus OB2 association, from the COBRaS Legacy programme. The radio observations potentially offer the most straightforward, least model-dependent, determinations of mass-loss rates, and can be used to help resolve current discrepancies in mass-loss rates via clumped and structured hot star winds. We report here that the 21 cm flux densities of O3 to O6 supergiant and giant stars are less than $\sim$ 70 $\mu$Jy. These fluxes may be translated to `smooth' wind mass-loss upper limits of $\sim$ {4.4 $-$ 4.8} $\times$ 10$^{-6}$ M$_\odot$ yr $^{-1}$ for O3 supergiants and $\lesssim$ {2.9} $\times$ 10$^{-6}$ M$_\odot$ yr $^{-1}$ for B0 to B1 supergiants. The first ever resolved 21 cm detections of the hypergiant (and LBV candidate) Cyg OB2 \#12 are discussed; for multiple observations separated by 14 days, we detect a $\sim$ {69}$\%$ increase in its flux density. Our constraints on the upper limits for the mass-loss rates of evolved OB stars in Cyg OB2 support the model that the inner wind region close to the stellar surface (where H$\alpha$ forms) is more clumped than the very extended geometric region sampled by our radio observations.}

\end{abstract}



\section{Introduction}

The Cyg OB2 Radio Survey (COBRaS) is an e-MERLIN Legacy Project to carry out a deep imaging radio survey of the central region of the Cygnus OB2 association (www.merlin.ac.uk/legacy/projects/cobras.html). The principal component of this project (252 hours) will be to map the core of Cyg OB2 at 5\,GHz (C-band; 6 cm; 2 GHz full bandwidth), going to a depth of $\sim$ 3\,$\mu$Jy (1-$\sigmaup$). Prior to these observations which are due in late 2016, additional pointings (42 hrs) at 1.4\,GHz (L-band; 21cm; 512 MHz full bandwidth) have been secured during 2014. We report here on results from the supplementary L-band datasets, and specifically on the constraints they provide on the mass-loss rates of OB stars and the nature of their outer wind regions.

Cygnus X is one of the richest star formation regions in the Galaxy. It hosts several OB associations, numerous young open clusters, tens of compact H\textsc{II} regions and star formation regions, a supernova remnant, and a superbubble blown by the collected stellar winds of the massive stars (e.g. \citealp{knodlseder_etal_2004}, \citealp{trapero_etal_1998}). At the core of Cygnus X is the Cyg OB2 association, which with a total cluster mass estimated to be $\sim$ 3$\times 10^4$\solmas, can be considered more as a massive cluster than an open OB association (\citealp{knodlseder_2000}; \citealp{wright_etal_2010}). The Cyg OB2 association is a uniquely important laboratory for studying the collective and individual properties of massive stars, and (possibly triggered) active star-formation. The stellar population of Cyg OB2 has been the focus of several studies across different wavebands (e.g. {\citealp{massey_thompson_1991}; \citealp{herrero_etal_2001}}; \citealp{comeron_etal_2002}; \citealp{setiagunawan_etal_2003}; \citealp{wright_etal_2014}; \citealp{rauw_etal_2015}). It has also been the target of radial velocity surveys (e.g. \citealp{kiminki_etal_2007}; \citealp{kobulnicky_etal_2012}). We lean here in particular on the recent census of \citet{wright_etal_2015} who list 169 OB stars, including 52 O-type and 8 normal early B supergiant. With an estimated cluster age of $\sim$ 2 Myr (\citealp{colombo_etal_2007}), Cyg OB2 is not only very rich in stellar density but also in its diversity. {The greater Cygnus X region includes} Be stars, many Young Stellar Objects (YSOs), two known Wolf-Rayet stars (WR 145, WR 146), two candidate Luminous Blue Variable (LBV) stars (G79.29+0.46, Cyg OB2 \#12), a red supergiant (IRC+40 427), a B[e] star (MWC 349), H\textsc{II} regions with groups of massive stars around them (DR 15, DR 18) and a gamma-ray source (TeV J2032+4130). Cyg OB2 is relatively close-by (at $\sim$ 1.4\,kpc), heavily obscured (as is the whole Cygnus X region), and located behind the Great Cygnus Rift. There is large and non-uniform visual extinction ranging from 4 to 10 mag (\citealp{knodlseder_2000}), thus making the association ideally studied at radio wavelengths.

Second only to the initial stellar mass, the mass-loss rates of massive stars determine the final stellar mass, and thereby, the type of compact stellar remnant for all stars more massive than about 8 \solmas. The amount of mass shed during main- and post-main-sequence evolution determines whether a star becomes a black hole or a neutron star and specifies the type of supernova or gamma-ray burst that it may produce. Knowledge of stellar mass-loss remains one of the most uncertain parameters in massive star evolution because of the unknown amount of clumping in the stellar winds. {Results since the late 20th century have} strongly challenged the canonical model of stellar-wind mass-loss in massive stars by emphasising uncertainties in small-scale clumping and large-scale structure in the outflows. {In normal OB-type stars, small-scale clumps are optically thin in H$\alpha$ and most likely also in radio emission} leading to derived mass-loss rate ($\rm{\dot{M}}$) diagnostics that {have been found to disagree with one another} by a factor of {2 to 10 (e.g. \citealp{drew_1990};} \citealp{puls_etal_2006}; \citealp{prinja_massa_2010}; \citealp{muijres_etal_2011}). The observations indicate that the winds universally contain large structures and small-scale clumping that are only partially characterised observationally, with effects on mass-loss rates {that have yet to be fully understood in conjunction with one another (see \citealt{sundqvist_etal_2014}).} There is thus a pivotal requirement to constrain wind clumping as a function of radial distance/velocity from the surface of the star and make comparisons to theoretical predictions in order to derive reliable mass-loss rates. 

OB stars emit radio radiation through (thermal) free-free emission, due to electron-ion interactions in their ionised wind. The considerable advantage of using free-free radio fluxes for determining mass-loss for massive stars is that, unlike H$\alpha$ and UV, the emission arises at large radii in the stellar wind, where the terminal velocity will have been reached. The interpretation of the radio fluxes is more straightforward therefore and is not strongly dependent on details of the velocity law, ionisation conditions\footnote{Care must be taken when considering the ionisation state of He in the outer wind regions {for it has been known to alter inferred mass-loss rates \citep[see e.g.][]{lamers_leitherer_1993}.}}, inner velocity field, or the photospheric profile. Furthermore, the greater geometric region and density squared dependence of the free-free flux makes the radio observations very sensitive to clumping in the wind. The radio measurements can be directly compared to other density-squared diagnostics such as H$\alpha$, which in turn permits constraints on the relative amount of wind clumping as a function of velocity (see e.g. \citealp{blomme_etal_2002}; \citealp{puls_etal_2006}).

We report here on first performance and science results from the COBRaS L-band (21 cm) Legacy data. We focus in this study on detection limits on thermal emission from suspected luminous single O and early B stars. The targets examined here are stars predicted to have the densest winds and highest mass-loss rates from inner-wind diagnostics such as H$\alpha$.

\section{Observations and image processing}

The core e-MERLIN L-band COBRaS legacy observations presented here were made over a three day period from April 25th - 27th 2014 with additional observations taken on April 11th 2014. The central approximately 15 square arcminutes of Cygnus OB2 were observed using seven overlapping pointings. The point source J2007+404 was used to perform cycled phase calibration scans during the observations. Each Cyg OB2 pointing was observed for two phase-target cycle scans before moving onto the next pointing. This process was repeated for the duration of the campaign in order to provide a good hour-angle coverage and to maintain as similar a uv-coverage for all pointings as possible. In total each pointing was observed for approximately five hours on source. The observations were made using full stokes parameters at a central frequency of {1.51}\,GHz using 512\,MHz bandwidth split over 8 Intermediate Frequencies (IFs) and 512 channels per IF. 

A large portion of the observable bandwidth suffers from contamination by Radio Frequency Interference (RFI). The data were edited using the RFI-mitigation software SERPent (\citealp{peck_fenech_2013}), a programme developed for e-MERLIN that utilises the {\sc{parseltongue}} scripting environment, as well as editing tasks within AIPS (Astronomical Image Processing System). In total, approximately 25-30\% of the data for each pointing were removed because of RFI alone.

Observations of the amplitude calibrator 3C286 were used to set the flux density scale. J2007+404 was used as a point source calibrator to determine the passbands and relative gains of the antennas. The data were phase calibrated and were weighted according to the relative sensitivity of each e-MERLIN antenna prior to imaging. The calibration was performed using standard procedures within AIPS and parts of the e-MERLIN pipeline (\citealp{argo_2014}). See Morford et al. (2016, in preparation) for further details.

The AIPS task IMAGR was used to produce a 512$\times$512 image of each of the sources within our sample. There were no strong sources around the outside of each image frame meaning the cleaning procedure was not affected by any external sidelobes. Each of these images was subsequently primary beam corrected using the standard AIPS task PBCOR to correct for the change in sensitivity over the primary beam.

\section{Single massive star sample selection}

To investigate the mass-loss rates in the L-band COBRaS data, we chose to limit our initial sample selection to stars expected to have the densest winds. Starting with the recent catalogue from \citet{wright_etal_2015}, we chose only stars that are within our field-of-view and are classified as either OI-OIII or BI stars. We further limited our sample to those stars that are known to be single or if they are within a binary system, have sufficient separation from their companion that any wind-wind interaction is negligible i.e. the expected emission is purely thermal. This assumption is crucial in both deriving radio mass-loss rates and investigating wind structure as explored by \citet{blomme_etal_2003}. The presence of one (or more) companion stars will facilitate the production of non-thermal (synchrotron) emission within the colliding wind region(s) {(e.g. as has been shown with Cyg OB2 \#8A; \citealt{blomme_etal_2010})}. Any non-thermal emission will contribute to the 21cm flux and cause an over-estimate of the object's mass-loss rate.

We present in Table \ref{measflux} our final target sample of nine massive stars. The sample includes two early O supergiant stars, two mid O giant stars, and five early B supergiant stars, one of which is a candidate LBV. Flux densities have been measured with the AIPS task \textsc{TVSTAT}. Where the source is not detected a 3\,$\sigma$ limit is quoted. The noise-level at the position of the source was measured using a 2$''$ diameter circle to determine the quoted limit. Figure 1 shows sub-images of some of the sample from the COBRaS L-band data. 


\begin{tabfonta}
\begin{table*}
\begin{center}
\caption[]{COBRaS L-band measured flux densities for our sample of stars. For Cyg \#7, 8C and 12 the $\rm{T_{eff}}$, $\rm{v_{\infty}}$, log$g$ and $\rm{M_{spec}}$ values are taken from the literature where a full nLTE analysis has been conducted in their derivation. {Typical errors on these stellar parameters are $\Delta \rm{T_{eff}} = \pm 500-1000$K. $\Delta \rm{v_{\infty}} = \pm 50 - 100$kms$^{-1}$, $\Delta$log$g = \pm 0.1 - 0.38$dex and $\rm{M_{spec}}$ is uncertain between 35$\%$ and 50$\%$.} For the remaining stars the parameters are adopted from standard spectral type values taken from the references listed, {note that parameters taken from references 4 and {5} were derived from calibrations}. Predicted \mdot values are calculated using the prescription from \citet{vink_etal_2001} {with a revised metallicity value, Z = 0.013 {following} \citet{asplund_etal_2009}. We denote our derived mass-loss rate as \mdot$_{max}$ since we have set $f_{cl}$ = 1 (see Sect. \ref{discussion} for further details).}}
\begin{tabular}{|c|c|c|c|c|c|c|c|c|c|c|c|c|}
\hline
RA & DEC & S58	& MT91	& Other	& Spectral & $\rm{T_{eff}}$ & $\rm{v_{\infty}}$ & log$g$ & $\rm{M_{spec}}$ & Flux Density & {\mdot$_{max}$} & Predicted \mdot \\ 
(J2000)&(J2000)&&&&Type&(K)&\kms&&$\rm{M_{\odot}}$&($\muup$Jy)&$10^{-6}$\,\solmasyr&$10^{-6}$\,\solmasyr\\
\hline
20 32 40.88 & 41 14 29.3  & 12 & 304 & - & B3.5Ia+ & $13700^{\,\bf{1}}$ & $400^{\,\bf{1}}$ & $1.70^{\,\bf{1}}$ & $110^{\,\bf{1}}$ & {1013$\pm$55} & {5.4}$\pm$1.4 & {24.5} \\
20 33 14.16 & 41 20 21.5  & 7  & 457 & - & O3If & $45800^{\,\bf{2}}$ & $3080^{\,\bf{3}}$ & $3.94^{\,\bf{2}}$ & $65^{\,\bf{2}}$ & $<$72 & $<${4.8} & {3.5} \\
20 33 18.02 & 41 18 31.0  & 8C & 483 & - & O5III & $41800^{\,\bf{2}}$ & $2650^{\,\bf{3}}$ & $3.74^{\,\bf{2}}$ & $49^{\,\bf{2}}$ & $<$71 & $<${4.1} & {1.9} \\
\hline
20 33 08.78 & 41 13 18.1  & 22 & 417 & - & O3If & $42551^{\,\bf{4}}$ & $3150^{\,\bf{6}}$ & $3.73^{\,\bf{4}}$ & $67^{\,\bf{4}}$ & $<$61 & $<${4.4} & {4.3} \\
20 33 14.84 & 41 18 41.4  & 8B & 462 & - & O6.5III & $35644^{\,\bf{4}}$ & $2545^{\,\bf{6}}$ & $3.63^{\,\bf{4}}$ & $34^{\,\bf{4}}$ & $<$78 & $<${4.3} & {0.7} \\
20 32 39.06 & 41 00 07.8  & - & - & E47 & B0Ia & $28100^{\,\bf{5}}$ & $1535^{\,\bf{6}}$ & $2.99^{\,\bf{5}}$ & $25^{\,\bf{5}}$ & $<$87 & $<${2.9} & {0.8} \\
20 33 39.14 & 41 19 26.1  & 19  & 601 & -  & B0Iab & $28900^{\,\bf{5}}$ & $1535^{\,\bf{6}}$ & $3.13^{\,\bf{5}}$ & $31^{\,\bf{5}}$ & $<$63 & $<${2.2} & {1.1} \\
20 33 30.81 & 41 15 22.7  & 18 & 556 & - & B1Ib & $21700^{\,\bf{5}}$ & $1065^{\,\bf{6}}$ & $2.67^{\,\bf{5}}$ & $22^{\,\bf{5}}$ & $<$73 & $<${1.8} & {1.6} \\
20 33 33.97 & 41 19 38.4  & - & 573 & - & B3I & $16400^{\,\bf{5}}$ & $590^{\,\bf{6}}$ & $2.16^{\,\bf{5}}$ & $19^{\,\bf{5}}$ & $<$58 & $<${0.8} & {1.4} \\

\hline
\end{tabular}
\label{measflux}
\footnotesize{References: {\bf 1} \citealp{clark_etal_2012}, {\bf 2} \citealp{mokiem_etal_2005}, {\bf 3} \citealp{herrero_etal_2001}, {\bf 4} \citealp{martins_etal_2005}, {\bf 5} \citealp{searle_etal_2008}, {\bf 6} \citealp{prinja_etal_1990}.}
\end{center}
\end{table*}
\end{tabfonta}

\section{Adopted fundamental parameters}

\label{adoptedparams}

Out of our target sample of nine OB stars, only one source is detected in the COBRaS L-band observations, the candidate LBV star Cyg OB2 \#12. For the remaining sources we use the observed flux density limits to calculate a smooth-wind mass-loss upper limit (listed in Table \ref{measflux}) by use of Equation 1, taken from \citet{wright_barlow_1975}.

\begin{equation}
\begin{aligned}
S_{\nu} = 23.2 \left( \frac{\dot{M}}{\mu v_{\infty}} \right)^{4/3} \frac{1}{D^2} \left(\gamma g_{ff} \nu \overline{Z^2} \right)^{2/3},
\end{aligned}
\end{equation}

where, S$_{\nu}$ is our observed radio flux in Jy measured at frequency $\nu$ in Hz; \mdot is in \solmasyr; v$_{\infty}$ is in kms$^{-1}$; D is the distance in kpc (see section \ref{distance} for further details). {Whilst hydrogen is expected to be fully ionised within the winds of OB stars, the ionisation state of helium depends on the stellar T$_{eff}$, the radial distance, and the wind density. Models created using the model atmosphere code CMFGEN \citep{hillier_miller_1998}, which simultaneously fit the multi-wavelength (UV through radio) observations of Cyg OB2 \#7 (e.g. \citealt{najarro_etal_2008, najarro_etal_2011} and {F. Najarro}, private communication) clearly suggest that He$^+$ dominates over He$^{2+}$ even in the most favourable scenario of vanishing clumping in the outer wind. In the case of residual clumping in the radio region (see Sect. \ref{discussion}), recombination would be enhanced, and therefore  He$^+$ would be even more dominant over He$^{2+}$. We adopt therefore that helium is singly ionised in the radio emitting region for all stars in our sample with the exception of Cyg OB2 \#12 for which helium is neutral in the radio formation region (\citealt{clark_etal_2012}, {F. Najarro}, priv. comm.). Furthermore, we assume a helium abundance of n$_{He}/$n$_{H}$ = 0.1 for all stars within our sample (see Section \ref{cyg7} in regards to Cyg OB2 \#7). The mean atomic weight of gas $\mu$, is taken to be 1.27 (1.4 in the case of Cyg OB2 \#12), the ratio of electron to ion density $\gamma$ = 1.0, and the mean ionic charge $\overline{Z^2}$ = 1.0. The gaunt factor,}

\begin{equation}
\begin{aligned}
g_{ff} \approx 9.77 \left( 1+0.13 \log \left(T_{e}^{3/2}/\nu \sqrt{(\overline{Z^2})}\, \right)\right),
\end{aligned}
\end{equation}

{as defined by \citet{leitherer_robert_1991},} has been calculated for each object assuming a constant relation between the stellar effective temperature and the electron temperature of the wind T$_e$ = 0.5T$_{eff}$ (\citealp{drew_1989}), where both T$_{eff}$ and T$_e$ are in K. {While the bound-free gaunt factor is known to play a role in this calculation, its contribution is found to be negligible (at $\lambda$ = 21cm) and has been subsequently disregarded (see Table IV of \citealt{waters_lamers_1984}).} 

In all cases, we have assumed that the radio emission is purely thermal. Additionally, we derive predicted mass-loss rates following the prescription from \citet{vink_etal_2001}, {utilising the revised solar metallicity values as given in \citet{asplund_etal_2009}.} For both Cyg OB2 \#7 and \#8C we adopt stellar and wind parameters from a full nLTE (non-Local Thermodynamic Equilibrium) analysis undertaken with the model atmosphere code FASTWIND by \citet{mokiem_etal_2005}. We note here that the adopted wind terminal velocity, $v_{\infty}$ values for these two O-type stars were first derived by \citet{herrero_etal_2001} in their study of HST STIS UV spectra. For Cyg OB2 \#12 we adopt parameters derived from \citet{clark_etal_2012}, who analysed UV-radio photometric and spectroscopic datasets using CMFGEN. For the rest of our target sample, no previous nLTE analysis had been conducted and as a result we rely on calibrated stellar parameters as a function of spectral type. For the O stars, we adopt those derived in \citet{martins_etal_2005} and for the B supergiant stars we refer to \citet{searle_etal_2008}.

\subsection{Distance to Cyg OB2}
\label{distance}

The observed free-free thermal radio emission in the winds of these hot stars is inversely proportional to the square of the distance to these objects. Unfortunately, the distance to the Cyg OB2 association is still uncertain, with estimates in the literature spanning between 0.9\,kpc and 2.1\,kpc. Using both spectroscopic and photometric observations to infer its distance becomes complicated since the region is known to suffer from variable extinction. Its location within the Galaxy also adds further complication since it lies at $l = 80^{\circ}$ where the relation between radial velocity and distance is poorly defined (see. \citealp{dame_thaddeus_1985}; \citealp{dame_etal_2001}). \citet{massey_thompson_1991} used spectroscopy and photometry of 63 stars in Cyg OB2 to infer a distance of 1.71\,kpc to the association, a value in good agreement with previous studies. The MK optical spectra of 14 Cyg OB2 stars were used by \citet{hanson_etal_2003} to derive a much closer distance estimate (D $\sim$ 1.45\,kpc). Subsequent work by \citet{negueruela_etal_2008} supported this revised distance in fitting model isochrones to a semi-observational HR diagram of Cyg OB2. Taking a different approach, \citet{linder_etal_2009} analysed the light curve of the eclipsing binary Cyg OB2 \#5, yielding a distance of 0.90$-$0.95 kpc. Whilst reporting that their result is in need of confirmation, they highlight the implications this distance estimate would have upon the stars within Cyg OB2, greatly reducing both the luminosities and mass-loss rate of its massive star members. More recently, measurements of the trigonomic parallaxes and proper motions of five star-forming regions within the Cygnus X complex gave a distance of $1.40 \pm 0.08$\,kpc to the region (\citealp{rygl_etal_2012}). Furthermore \citet{kiminki_etal_2015} measured the distance to four eclipsing binary members of Cyg OB2, obtaining a weighted average distance of $1.33 \pm 0.06$\,kpc. This result sits slightly lower than those obtained via spectro-photometric methods, although interestingly \citet{hanson_etal_2003} noted that if they were to adopt the cooler $T_{eff}$ scales of \citet{martins_etal_2002, martins_etal_2005}, their distance estimate would reduce to 1.2\,kpc. This was rejected by \citet{hanson_etal_2003} as moving the association closer would reduce the luminosity of the association's supergiants and hence also reduce their mass-loss rate, making them more discrepant than those predicted from stellar wind theory (e.g. \citealp{vink_etal_2001}). Growing evidence exists {that may require the revision of} accepted mass-loss rates downwards (see. \citealp{puls_etal_2008}) and indeed this uncertainty is the motivation behind this work. For this study, we adopt a distance to Cyg OB2 of $1.4 \pm 0.1$\,kpc. {Where mass-loss rates have been compared to those taken from the literature, we also provide values scaled to our adopted distance of 1.4 kpc}.

\section{Mass-loss rates in the radio regime}


\subsection{Cygnus OB2 \#7}
\label{cyg7}

This O3If star, has previously been extensively observed and modelled with nLTE codes such as FASTWIND and CMFGEN (see \citealp{herrero_etal_2000, herrero_etal_2002, herrero_etal_2003}; \citealp{mokiem_etal_2005}; \citealp{puls_etal_2006}; \citealp{najarro_etal_2011}; \citealp{maryeva_etal_2012, maryeva_etal_2013}). Despite being part of previous radio surveys (\citealp{bieging_etal_1989}; \citealp{setiagunawan_etal_2003}) it is yet to be detected at radio wavelengths; with a previous 3\,$\sigmaup$ flux limit of $1.0 $ mJy at L-band (21cm) \citep{setiagunawan_etal_2003}.

Interestingly, despite vastly improved sensitivities with e-MERLIN, the COBRaS L-band observations provide only an upper limit for Cyg OB2 \#7. The observations reported here improve upon previous L-band (21cm) upper limit flux densities by a over a factor of $\sim$10, reaching a 3\,$\sigmaup$ flux density of 72 $\mu$Jy. This corresponds to an upper limit on the mass-loss rate of Cyg OB2 \#7 of {4.8}$\times 10^{-6}$\,\solmasyr. In comparison to the most recent values obtained via nLTE analysis of the H$\alpha$ line, our result is approximately a factor of two lower (e.g. \citealp{maryeva_etal_2013}). Using standard stellar parameters as a function of spectral type (see \citealp{martins_etal_2005}) and the recommended mass-loss recipe from {\citet{vink_etal_2001}} to calculate a predicted \mdot for Cyg OB2 \#7, provides values in good agreement with our result (see Table. \ref{measflux}).  

The first full nLTE analysis was carried out on Cyg OB2 \#7 (\citealp{herrero_etal_2000}), using the model atmosphere code FASTWIND. {H$\alpha$ and HeII$\lambda$4686 in combination with other optical lines were used to sample the inner-most wind regions (R $ \lesssim$ 3 {R$_{\star}$; see e.g. \citealt{prinja_etal_1996}}) to} derive a value of \mdot = 11.2$\times 10^{-6}$\,\solmasyr {({8.3}$\times 10^{-6}$\,\solmasyr at 1.4kpc)}. {Furthermore, the same authors used an updated version of the FASTWIND code to account for both metal-line blocking and blanketing, that led to the derivation of similar \mdot values (\citealp{herrero_etal_2002,herrero_etal_2003}).} \citet{mokiem_etal_2005} combined the FASTWIND code with a genetic algorithm based optimisation routine known as PIKAIA to automate the spectrum fitting process and again derived a `smooth wind' mass-loss rate of $\sim $10$ \times 10^{-6}$\,\solmasyr {({7.4}$\times 10^{-6}$\,\solmasyr at 1.4kpc)}.

Subsequent studies of Cyg OB2 \#7 were carried out by \citet{puls_etal_2006}, who conducted a multi-wavelength analysis incorporating radio, infrared (IR) and H$\alpha$ observations. In doing so, they probe different wind regions to put constraints on the radial stratification of the clumping factor. Their multi-wavelength approach, constrained by non-detections in VLA observations at 6 and 3.5 cm, derived an upper limit on \mdot comparable to that found here. {Note that \citet{puls_etal_2006} assumed a Helium enrichment $Y_{He}$ = 0.21, contrary to the value used here of $Y_{He}$ = 0.1. Assuming the free-free thermal flux scales with frequency as $\nu^{0.6}$, the 21cm upper limit found here can be translated into a flux of 145 and 202 $\mu$Jy at 6 and 3.5 cm respectively, i.e. consistent with those derived by \citet{puls_etal_2006}. The same authors fixed the outer-wind clumping factor $f_{cl}$ = 1 to derive a maximum mass-loss for Cyg OB2 \#7. Whilst the results found here support this notion and provide good evidence that the outer wind regions are less clumped than the inner (H$\alpha$) wind regions, we stress that the clumping factor cannot be fully constrained at the radio photosphere without explicit knowledge of the star's mass-loss rate.} 

Further nLTE analysis utilising the H$\alpha$ line diagnostic (see \citealp{najarro_etal_2011}; \citealp{maryeva_etal_2012, maryeva_etal_2013}), all derive a `smooth wind' \mdot value consistent with previous nLTE analysis of the star (i.e. are around a factor of two larger than found here). In the case of Cyg OB2 \#7, the discrepancy between inner-wind region (H$\alpha$) and outer-wind region (radio) mass-loss rates is clear and must be attributed to the effect of wind structure. We return to this discussion in Sect. \ref{discussion}.

\subsection{Cygnus OB2 \#8C}
\label{cyg8c}

Cyg OB2 \#8C was re-classified as an OIII by \citet{kiminki_etal_2007} though it had previously been considered to have an OIf spectral type (\citealp{massey_thompson_1991}). This star has been observed as part of previous radio surveys (e.g. \citealp{setiagunawan_etal_2003}), though it has not been detected. These COBRaS data provide the most sensitive radio observations of Cygnus OB2 to date and give a 3\,$\sigmaup$ upper limit to the flux density of {71\,\mujy}, corresponding to an \mdot upper limit of {4.1}$\times 10^{-6}$ \solmasyr at 21cm.

Line synthesis modelling of this star has previously been used to measure a mass-loss rate from H$\rm{\alpha}$ profiles. \citet{herrero_etal_2002} calculate a \mdot of 2.3$\times 10^{-6}$\,\solmasyr {({1.7}$\times 10^{-6}$\,\solmasyr at 1.4kpc)} and later a lower value of 1.7$\times 10^{-6}$\,\solmasyr \citep[{{1.3}$\times 10^{-6}$\,\solmasyr at 1.4kpc;}][]{herrero_etal_2003}. Similarly, \citet{mokiem_etal_2005} used FASTWIND to calculate a mass-loss rate of 3.4$\times 10^{-6}$\,\solmasyr {({2.5}$\times 10^{-6}$\,\solmasyr at 1.4kpc)}. \citet{puls_etal_2006} utilised a 200\,\mujy radio upper limit taken from \citet{bieging_etal_1989} in their calculation. Though their analysis of Cyg OB2 \#8C failed to fully constrain the mass-loss and clumping properties across all regions they quote a \mdot upper limit of 4.3$\times 10^{-6}$\,\solmasyr {({3.2}$\times 10^{-6}$\,\solmasyr at 1.4kpc)} based primarily on the radio flux density upper limit. However, they also calculate a lower H$\alphaup$ derived \mdot of 3.5$\times 10^{-6}$\,\solmasyr {({2.6}$\times 10^{-6}$\,\solmasyr at 1.4kpc)}.

\begin{figure*}
\begin{center}$
\begin{array}{cc}
\includegraphics[width=0.4\textwidth]{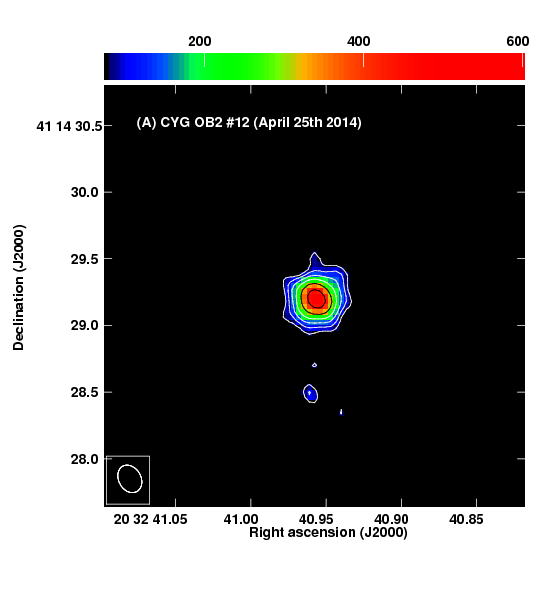} &
\includegraphics[width=0.4\textwidth]{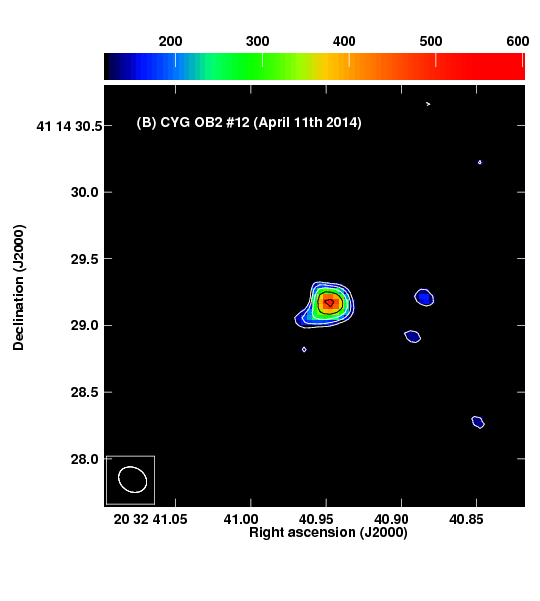} \\
\includegraphics[width=0.4\textwidth]{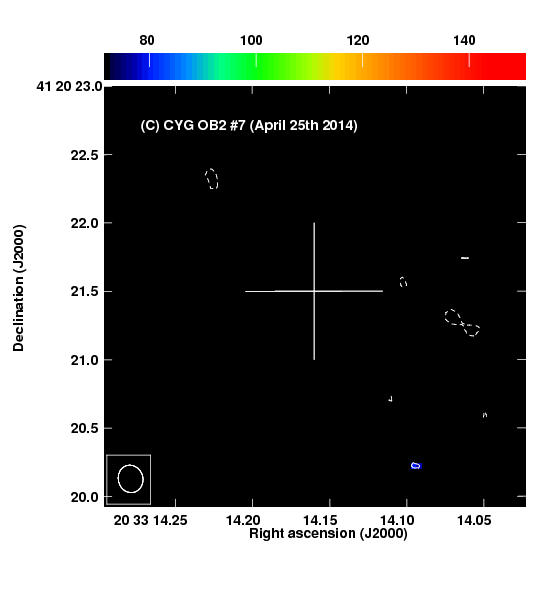} &
\includegraphics[width=0.4\textwidth]{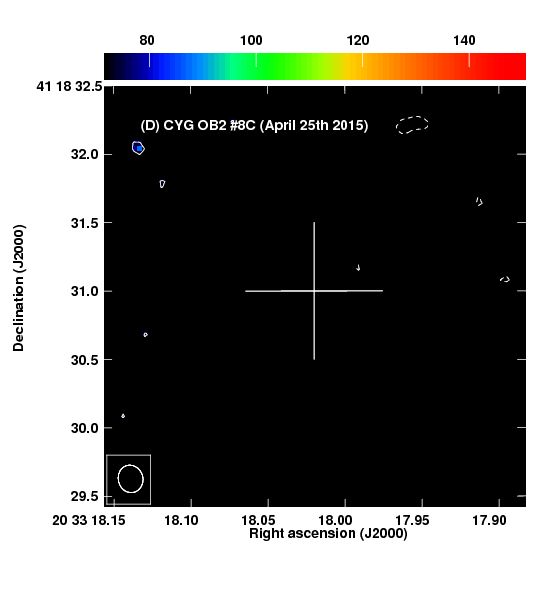}
\end{array}$
\caption{Images of three of the nine target sample stars from the COBRaS 21cm Legacy observations, images A, C and D are all from observations taken between the 25th and 27th of April: (A) Cyg OB2 \#12 the first ever resolved image at 21cm, 1$\sigma$ rms $=$ 24 $\mu$Jy$/$beam; (B) a second image of Cyg OB2 \#12 from observations taken on April 11th 2014, 1$\sigma$ rms $=$ 40 $\mu$Jy$/$beam; (C) a blank field image of Cyg OB2 \#7, image 1$\sigma$ rms $=$ 24 $\mu$Jy$/$beam; (D) a blank field image of Cyg OB2 \#8C, image 1$\sigma$ rms $=$ 24 $\mu$Jy$/$beam. Upper horizontal bar displays the colour scale of each pixel in units of $\mu$Jy$/$beam, all contour levels are {-1, 1, 1.4, 2, 2.8, 4, 5.7, 8, 11.3, 16} $\times$ 3$\sigma$ image rms.}
\label{cyg12_25th}
\end{center}
\end{figure*}

\subsection{Cygnus OB2 \#12}

Cyg OB2 \#12 is a luminosity class Ia$+$ star \citep{keenan_1971}. \citet{vanGenderen_1982} later described this class as `Blue Hypergiants' (BHG) which not only differ from `Blue Supergiants' (BSG) by their large luminosity, but also spectroscopically with the presence of P Cygni balmer line emission. Cyg OB2 \#12 has always been a particularly interesting BHG case due to its extremely high luminosity \citep{schulte_1958}. Whilst showcasing some properties of a LBV star, it is also missing some of their typical characteristics leading to uncertainties upon its exact classification (\citealp{clark_larionov_2005}; \citealp{clark_etal_2012}).

LBVs are massive, unstable stars found in the upper left hand region of the HR diagram. Whilst having extremely high luminosities ($\sim$10$^6$ L$_{\odot}$) and large mass-loss rates (up to 10$^{-4}$ M$_{\odot}$yr$^{-1}$), they are found to be significantly variable both photometrically and spectroscopically \citep{humphreys_davidson_1994}. They have been observed to have two types of variability, the first of which is reflected in their visual magnitude and is a result of how they cool and expand (heat and contract), shifting to redder (bluer) colours. The second is a consequence of significant mass-loss episodes such as the case for $\eta$ Carinae. These eruptions are far rarer, with only two known examples in our Galaxy \citep{clark_larionov_2005}.

Here, we report on the first ever {resolved detection at 21cm} of Cyg OB2 \#12. From the COBRaS L-band observations taken between the 25th and 27th of April 2014, we observe a flux density for Cyg OB2 \#12 of {1013 $\pm$ 55} \,\mujy (see Figure. \ref{cyg12_25th}A). Assuming a smooth wind model ({$f_{cl}$ = 1}) and that the flux received is purely thermal free-free emission, we calculate its mass-loss rate to be \mdot $=$ {5.4} $\pm 1.4 \times 10^{-6}$\,\solmasyr. {We note here our assumption that Hydrogen is still ionised in the outer wind regions despite the cool temperature of Cyg OB2 \#12.} \citet{clark_etal_2012} modelled Cyg OB2 \#12 using CMFGEN to infer \mdot $= 3.0 \times 10^{-6}$\,\solmasyr {({2.1}$\times 10^{-6}$\,\solmasyr at 1.4kpc)}, with a clumping factor f$_{cl} = 25$. {This value of $f_{cl}$ was derived using a modified version of CMFGENs clumping prescription to account for the low terminal velocity of Cyg OB2 \#12. Predominantly constrained using the H$\alpha - \beta$ and Br$\alpha$ emission components, the IR, sub-mm and radio continuum \citep{clark_etal_2012}, this value of $f_{cl}$ = 25 holds from R $>$ 40 R$_{\star}$ ({F. Najarro}, priv. comm.) covering the entire radio emitting region. The derived \mdot translates into a an `unclumped' (smooth-wind) value of 15}$\times 10^{-6}$\,\solmasyr {({10.7}$\times 10^{-6}$\,\solmasyr at 1.4kpc;} see also Sect. \ref{discussion}), giving a discrepancy of a factor of $\sim$ {2} in comparison to our value. This alone highlights the uncertainty in current \mdot {diagnostics} and adds to the growing evidence for the disparity between different mass-loss diagnostics.

The uncertainty of this object and the ongoing debate upon its precise classification cannot be overlooked. The significantly lower than previously found \mdot value derived here could instead be explained by the variability of the object. COBRaS L-band observations were also obtained on the 11th of April 2014, some 14 days prior to the core of the observations presented here. Despite the relatively small time window between observations, we searched for any variability of the flux of Cyg OB2 \#12. With a sensitivity of $\sim$ 40 $\mu$Jy, the observations taken on the April 11th 2014 yielded a flux of {598 $\pm$ 61 $\mu$Jy (see Figure \ref{cyg12_25th}B) corresponding to a `smooth-wind' mass-loss rate of {3.6}$\times 10^{-6}$\,\solmasyr. We therefore observe a {50\% increase in the mass-loss rate of Cyg OB2 \#12 (or a 69\% increase in the flux density)} over the 14 day period. A possible constraint on the origin of this variation may be derived from considering the `effective radius' of the radio emission, defined as the radial distance at which the free-free optical depth is 0.244 \citep{wright_barlow_1975}. Using the stellar parameters found in Table \ref{measflux} {(and R$_{\star}$ = 246 R$_{\odot}$; \citealt{clark_etal_2012})}, the effective radius of 21cm emission in Cyg OB2 \#12 is {$\sim$ 86 R$_{\star}$}. To cover this distance at a constant velocity of v$_\infty$ = 400 kms$^{-1}$ would require $\sim$ 424 days, which is much longer than the 14 days between our two observation epochs. For comparison a typical O-type star with v$_\infty$ = 2600 kms$^{-1}$, T$_{eff}$ = 40 kK{, R$_{\star}$ = 10 R$_{\odot}$} and a mass-loss rate of 4 $\times 10^{-6}$ \solmasyr would need $\sim$ 9 days to cover the distance of its 21cm effective radius {($\sim$ 304 R$_{\star}$)}. We conclude that the variation in 21cm flux of Cyg OB2 \#12 cannot be due to a global mass-flux variation.} We note that the April 25th observations are approximately 3$\times$ the on-source integration time than those of the April 11th and the difference between the coverage in hour angle results in a different primary beam shape and size between the two observation epochs. The object is resolved in both epochs, with a deconvolved angular size of 285 milliarcseconds (mas) and 453 mas for the 11th and 25th of April observations respectively. Short term variations in the flux of Cyg OB2 \#12 have also been found in previous radio observations. \citet{bieging_etal_1989} found a 70\% variation {at 6\,cm} in the flux of Cyg OB2 \#12, whilst \citet{scuderi_etal_1998} observed a 50\% variation in its radio flux over the timescale of a month {at 2, 3.6 and 6\,cm}. Furthermore, its X-ray flux has also been found to vary on the 10\% level over timescales of up to a week \citep{rauw_2011}.

Cyg OB2 \#12 has been extensively studied. \citet{clark_etal_2012} report that the combination of its extremely high luminosity and low temperature imply its position on the HR diagram cannot be matched to any theoretical isochrone applied to its host association Cyg OB2. \citet{cazorla_etal_2014} looked at XMM-Newton and Swift X-ray observations of Cyg \#12 and find a marked decrease in X-ray flux in recent years (40\% from 2004 to 2011), compatible either with a wind-wind collision in a wide binary or the aftermath of a recent eruption. No evidence for a companion star had previously been found until \citet{caballeronieves_etal_2014} detected a close companion separated by an angular distance of $63.6$ mas. Their finding has since been confirmed by \citet{maryeva_etal_2016}, who further resolved a very faint third counterpart. Note however that \citet{caballeronieves_etal_2014} report that their detected secondary is too faint to substantially decrease the luminosity of Cyg OB2 \#12, and hence alter the conclusions drawn from \citet{clark_etal_2012}. Depending on the nature of Cyg OB2 \#12's companions, this potentially undermines the assumption that the radio flux detected in the COBRaS observations presented here is completely thermal in origin. However, it is important to note that any non-thermal emission as a result of a colliding wind region would only further contribute to the 21cm flux received here, implying a smaller contribution from thermal free-free emission. Hence the presence of non-thermal emission would only seek to lower the mass-loss rate derived here, in contradiction to previous \mdot estimates (e.g. \citealp{clark_etal_2012}).

\subsection{The remaining sample selection}

For the majority of the stars in our sample selection (i.e. those excluding Cyg OB2 \#7, Cyg OB2 \#8C and Cyg OB2 \#12) there is currently no individual nLTE modelling in the literature. For these stars we can only compare in Table 1 the mass-loss rate limits provided by this study with those predicted (as a function of spectral type), by the prescription from \citet{vink_etal_2001}. In general, there is a {broad} agreement between those calculated from the COBRaS L-band upper limits and those predicted for both the O and B stars within the sample. {Furthermore, \citet{puls_etal_2006} derived mass-loss rates consistent with the those calculated from the \citet{vink_etal_2001} recipe (despite using an `older' metallicity value of Z = 0.019 in the \mdot prescription) yet other diagnostics (across multiple wavebands) derive values 2-3 (with translates to $\sim$1.5-2 with Z = 0.013) times lower than these theoretical values (e.g. see \citealt{najarro_etal_2011}: IR lines; \citealt{cohen_etal_2014}: X-ray line emission; \citealt{sundqvist_etal_2011, surlan_etal_2013, sundqvist_etal_2014}: UV-lines including velocity porosity and optical lines). If we therefore assume that the theoretical prescription calculated here (with Z = 0.013) consistently gives \mdot values a factor of 1.5-2 times too large, we could postulate a value of $f_{cl}$ = 4 in the outer wind regions in order to pull our radio inferred mass-loss rates down by a factor of $\sim$2 to coincide with the {above discrepancy} found between the prescription from \citet{vink_etal_2001} and other diagnostics.} 

The \mdot upper limits of the two luminous O supergiant stars are broadly in agreement with the alternative \mdot predictions given by \citet{muijres_etal_2012}. {Furthermore, the \mdot upper limits of the two B supergiant stars Cyg OB2 \#12 and MT573 are found to lie underneath the predictions inferred from \citet{vink_etal_2001} by factors of approximately 5 and 2 respectively. These are the only two stars in our sample for which this is the case and interestingly the only objects in our sample who's T$_{eff}$ lies below the observed bi-stability jump at 20000 K {(e.g. \citealt{evans_etal_2004, crowther_etal_2006, markova_puls_2008}; see also \citealt{petrov_etal_2016} for the corresponding theoretical findings)}. Due to the lack of radio observations of B supergiant stars, this result could provide crucial information to the on-going debate regarding the increase in \mdot as a star crosses this bi-stability jump (e.g. see \citealt{vink_etal_1999, markova_puls_2008, petrov_etal_2016}).} We {feel it important to} note however, the large uncertainty upon \mdot predictions due to the dependence on the stellar mass which in turn may be uncertain up to 50\% (\citealp{martins_etal_2005}).

\subsection{A11}

A11 (or MT267), is an O-type star within our field-of-view. Listed in the catalogue of \citet{wright_etal_2015} as a single star of spectral type O7.5III, it initially met our target selection criteria. \citet{kobulnicky_etal_2012} however, show A11 to be a binary system with an O7.5III-I primary and a period of 15.511$\pm$0.056 days. {As such, this star was rejected by our selection criteria and subsequently not included in Table \ref{measflux}.} Furthermore, the variable H$\alphaup$ emission and observed x-ray variability suggests this to be an interacting binary. We detect A11 in the COBRaS L-band data with a flux density of 161 $\pm$ 27 $\mu$Jy. This result in comparison to the sample star flux densities in Table. \ref{measflux}, supports the notion that A11 is an interacting binary with non-thermal emission from a wind-wind collision region.

\section{Discussion}
\label{discussion}

We have used e-MERLIN observations of Cyg OB2 from the ongoing COBRaS project to demonstrate that the 21 cm flux densities of a sample of luminous, early O-type stars are below $\sim$ 70 $\mu$Jy. Under the assumption that the emission is entirely thermal in origin, and the stellar wind region is unclumped, we place upper limits of $\sim$ {4.4 $-$ 4.8} $\times$ 10$^{-6}$ M$_\odot$ yr $^{-1}$ on the mass-loss rates of O3 I stars; i.e. the  hottest and most luminous stars in our sample. The mass-loss rates of early B supergiants (B0 to B1) are constrained to less than $\sim$ {1.8 $-$ 2.9} $\times$ 10$^{-6}$ M$_\odot$ yr $^{-1}$. Adopting spectroscopic masses, our upper limits are broadly consistent with mass-loss rates derived from the semi-empirical prescriptions of \citet{vink_etal_2000, vink_etal_2001}{, with the exception of the LBV candidate Cyg OB2 \#12.} For luminous O stars the \citealt{vink_etal_2001} values are in turn consistent with the refined predictions of \citealt{muijres_etal_2012}, who solve the wind dynamics numerically. 

The O3 to O5 stars in our sample have an effective photospheric radius of more than 150 R$_\star$ at 21 cm and our observations thus sample the most outer regions of the stellar winds. Assumptions that the wind is unclumped or very weakly clumped in this region is essentially untested observationally. Given that free-free emission depends on density squared, our fluxes either correspond to a smooth wind mass-loss rate, or a lower mass-loss rate $\times$ $\sqrt{f_{cl}}$, where $f_{cl}$ is the clumping factor. Comparing to the primarily recombination-formed line-synthesis analyses (i.e. H$\alpha$, He{\sc II}) of Cyg OB2 \#7, which samples the inner-most wind regions (below $\sim$ 3 R$_\star$), \citet{herrero_etal_2002}, \citet{mokiem_etal_2005}, \citet{repolust_etal_2005}, \citet{maryeva_etal_2013} all derive a 'smooth wind' mass-loss rate of $\sim$ 8.0 $-$ 10 $\times$ 10$^{-6}$ M$_\odot$ yr $^{-1}$. These consistently high mass-loss rates can only be reconciled with our 21 cm {\it upper limit} of {4.8} $\times$ 10$^{-6}$ \solmasyr if the inner wind H$\alpha$ region (close to the stellar surface) is substantially more clumped than the radio free-free formation region sampled in our study. This result is in agreement with the clumped wind models discussed by \citet{puls_etal_2006}, and with the notion that there is a radial stratification of the clumping factor in the stellar winds of OB stars. However, the derived clumping factor (and therefore mass-loss rate) is dependent on the assumption adopted for the degree of clumping in the radio formation region.

Regarding the issue of structure in the outermost wind regions, the growth of the intrinsic line-deshadowing instability (LDI) has been numerically modelled by e.g. \citet{owocki_etal_1988}; \citet{feldmeier_1995}; \citet{dessart_owocki_2005}. The simulations show that the LDI leads to high-speed rarefactions that provide a basis for our interpretation of wind clumping. In their 1-d time-dependent hydrodynamical study of stochastic structure, Runacres {\&} Owocki (2002) model the evolution of clumped structure far from the stellar surface. Their models predict a rise in the clumping factor from the inner wind to $\sim$ 50 R$_\star$, and a subsequent decrease in the clumping factor to a residual value beyond $\sim$ 100 R$_\star$. Depending on the details, simulations predict that the stellar winds remain clumped deep into the radio formation region, with clumping factors between 2.5 to 6. As noted above, the single epoch radio continuum observations do not provide any direct information as to whether or not the OB stars winds are clumped beyond $\sim$ 100 R$_\star$.

The substantial 6 cm (C-band) e-MERLIN COBRaS Legacy observations, scheduled from October 2016 onward, will provide flux densities down to a 3$\sigma$ limit of $\sim$ 10 $\mu$Jy. These data will ultimately lead to the tightest constraints on the outer wind mass-loss rates of OB stars in Cyg OB2 for a wide range of effective temperature, luminosity and wind density.

\section*{Acknowledgments}

e-MERLIN is a national facility operated by The University of Manchester on behalf of the Science and Technology Facilities Council (STFC). {\sc{parseltongue}} was developed in the context of the ALBUS project, which has benefited from research funding from the European Community's sixth Framework Programme under RadioNet R113CT 2003 5058187. J. Morford and D. Fenech wish to acknowledge funding from an STFC studentship and STFC consolidated grant (ST/M001334/1) respectively. We thank Paco Najarro and Ian Stevens for useful discussions. We are also grateful to the referee Jo Puls for his input and suggestions on the original manuscript.

\bibliographystyle{mn2e}
\bibliography{lb_pap1_ref}

\label{lastpage}

\end{document}